\documentclass[conference,onecolumn]{IEEEtran}
\IEEEoverridecommandlockouts
\usepackage{cite}
\usepackage{amsmath,amssymb,amsfonts}
\usepackage{algorithmic}
\usepackage{graphicx}
\usepackage{textcomp}
\usepackage{xcolor}
\usepackage{float}
\usepackage{enumerate}
\def\BibTeX{{\rm B\kern-.05em{\sc i\kern-.025em b}\kern-.08em
    T\kern-.1667em\lower.7ex\hbox{E}\kern-.125emX}}
\begin{document}

\title{BER Analysis of Reconfigurable Intelligent Surface Assisted Downlink Power Domain NOMA System
}

\author{\IEEEauthorblockN{Vetrivel Chelian Thirumavalavan}
\IEEEauthorblockA{\textit{Research Scholar, Dept. of ECE} \\
\textit{Thiagarajar College of Engineering}\\
Madurai, India \\
vetrivelchelian@student.tce.edu}
\and
\IEEEauthorblockN{Thiruvengadam S Jayaraman}
\IEEEauthorblockA{\textit{Professor, Dept. of ECE} \\
\textit{Thiagarajar College of Engineering}\\
Madurai, India \\
sjtece@tce.edu}

}

\maketitle

\begin{abstract}
The use of software controlled passive Reconfigurable Intelligent Surface (RIS) in wireless communications has attracted many researchers in recent years. RIS has a certain degree of control over the scattering and reflection characteristics of the electromagnetic waves, compared to the conventional communications in which the received signal is degraded due to the uncontrollable scattering of the transmitted signal and its interaction with the objects in propagating medium. Further, in RIS assisted communications, the phases of the multiple incoming signals can be controlled to enable constructive addition of multiple signals from different channel paths to improve Signal to Noise Ratio (SNR). On the other hand, Non-Orthogonal Multiple Access (NOMA) provides massive connectivity and low latency. The power domain variant NOMA uses superposition coded symbols with different powers for different user symbols. In this paper, a novel RIS assisted downlink NOMA system is proposed by combining the merits of both RIS and NOMA to improve the reliability of the system. Analytical expressions are derived for the Bit Error Rate (BER) performance of the proposed RIS assisted power domain NOMA system. The BER performance of the proposed system is analyzed using the numerical simulation results. It is observed that the proposed system has better performance than the conventional NOMA system.
\end{abstract}


\section{Introduction }
In practice, when multiple signals from different paths or channels arrive at the receiver in same phase, the total signal at the receiver will experience significant signal strength improvement due to the constructive addition. An early idea to exploit the phases of the multiple incoming signals is through diversity reception techniques such as Maximum Ratio Combining (MRC) receiver. Instead of modifying the transceivers, Reconfigurable Intelligent Surfaces (RIS) can modify the signal through wireless channel. Utilization of RIS for assisting wireless communication has numerous benefits. One of the benefits is the facilitation of ultra-reliable wireless communication even at very low Signal to Noise Ratios (SNR). Few others benefits include low cost implementation and energy efficient hardware. RIS metasurfaces can be fabricated to have the ability to reconfigure its reflection properties of the signal through Micro-Electronic Mechanical Systems (MEMS) or varactor diodes \cite{tan2016increasing}. The working principle of RIS is simple. Arbitrary incident waves are phase tuned by the discrete reflecting elements and the reflected waves out of RIS are deterministic and coherent \cite{zhang2018space}. This way operators can have a certain control over the random channel behavior. Basar derived the analytical expressions of symbol/bit error probability for RIS assisted wireless communications in \cite{maina}. 

Non Orthogonal Multiple Access (NOMA) technology shares the same resource elements in time, frequency, space and code domain. Instead, users are separated in power domain and its a strong contender for multiple access technology in future wireless systems \cite{kayakara}.

 In this paper, a novel RIS assisted downlink NOMA system is proposed by combining the merits of both RIS and NOMA to improve the reliability of the system. In the proposed system, the NOMA downlink is considered from Base station (BS) to two different users, Near User (NU) and Far User (FU) with the communication taking place through RIS plane. It is assumed that there is no Line of Sight (LoS) link between BS and any of the users. Error performance of QPSK modulated NU and BPSK modulated FU are derived and simulated. The derivation can be easily extended to other M-PSK or M-QAM systems, it also can be extended to any number of users. Error performance of NU and FU in the proposed system is analyzed for different number of reflecting elements. An unequal RIS element allocation schemes for NU and FU are also investigated.

\section{Proposed RIS Assisted NOMA}

\begin{figure}
\centering
\includegraphics[scale=0.7]{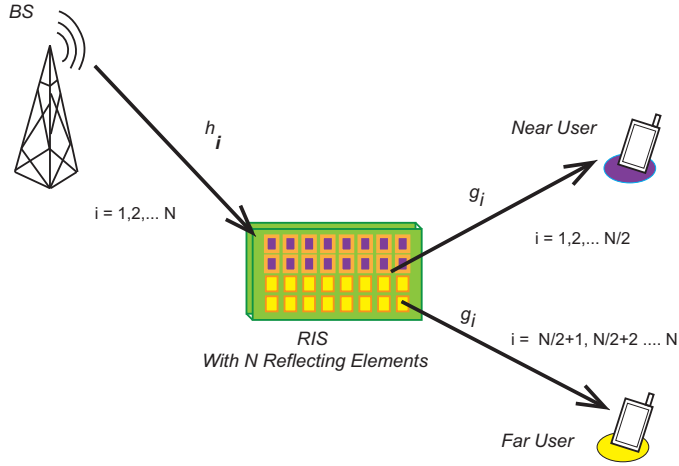}
\caption{Proposed RIS Assisted Downlink Power Domain NOMA System}
\label{LISassisted2ray}
\end{figure}

The system model for the proposed RIS assisted downlink power domain NOMA is given in Fig.\ref{LISassisted2ray}. $N$ is the number of reflecting elements in the RIS. It is assumed that $ N_{nu}$ and $N_{fu} $ are the number of reflectors allocated to NU and FU. As RIS elements can be independently configured to desired angle of departure, it is assumed that only the signal from allocated elements reach the respective users NU and FU. Sum of $ N_{nu}$ and $N_{fu} $ is equal to $N$. If equal ($ N_{nu} = N_{fu} = N/2 $) number of reflecting elements are allocated, the received signal at NU, ${r}_{nu}$ is given as

\begin{equation}
{r}_{nu}=\left [\mathbf{g}_{nu}^{T} \mathbf{\Phi h}_{nu} \right ] x + n_{nu}
\end{equation}
where $\mathbf{h}_{nu}=\left[ h_{1}~ h_{2}~ \ldots h_{N/2} ~\right]^{{T}}$, $\mathbf{g}=\left[ g_{1}~ g_{2}~ \ldots~ g_{N/2} \right]^{{T}}$, $\Phi = diag\left ( \left [e^{-j \phi_{1}}~ e^{-j \phi_{2}} \ldots e^{-j \phi_{N/2}}\right ] \right )$. $h_i$ is the single antenna fading channel between BS and RIS $i^{th}$ reflecting surfaces and $g_i$ is the fading channel between the RIS $i^{th}$ reflecting surfaces to a single-antenna NU. $h_{i}$ and $g_{i}$ are Rayleigh distributed Random Variables (RV) and are independent to each other. They are defined as ${h}_i= \alpha_{i}e^{\theta_{i}}$ and  ${g}_i= \beta_{i}e^{\psi_{i}}$. Individual mean and variances for each $\alpha_{i}$ and $\beta_{i}$ are assumed to be
$E[x] = \sigma \sqrt{\frac{\pi}{2}}$, $VAR[x] = \sigma^{2} \frac{(4 - \pi)}{2}$. $\phi_i$ is the adjustable phase induced by the $i^{th}$ reflecting meta-surface of the RIS. $n_{nu}$ is Additive White Gaussian Noise (AWGN) with zero-mean and variance $No/2$ 

From ~\cite{Nomaa}, the superposition coded symbol $x$ is given by
\begin{equation}
x =\sqrt{\varepsilon_1}x_{1} ~+~\sqrt{\varepsilon_2}x_{2} 
\label{super45}
\end{equation}
 where $\varepsilon_{1}$ is the NU energy and $\varepsilon_{2}$ is energy allocated for FU. $E_s$ is the superposition symbol power, we assume $\varepsilon_{1}=\alpha E_s$ and $\varepsilon_{2}=(1-\alpha) E_s$ and $\alpha$ is the NOMA superposition coding power allocation coefficient.

Similarly, for the FU received signal is given by

\begin{equation}
{r}_{fu}=\left [\mathbf{g}_{fu}^{T} \mathbf{\Phi}_{fu} \mathbf{h}_{fu} \right ] x + n_{fu}
\end{equation}

where $\mathbf{h}_{fu}=\left[ h_{{N/2}+1} ~ h_{{N/2}+2}~ \ldots h_{N} ~\right]^{{T}}$, 
$\mathbf{g}_fu=\left[ g_{{N/2}+1} ~ g_{{N/2}+2}~ \ldots g_{N} \right]^{{T}}$,
 $\Phi_{fu} = diag\left ( \left [e^{-j \phi_{{N/2}+1}}~ e^{-j \phi_{{N/2}+2}} \ldots e^{-j \phi_{N}} \right ] \right )$. $g_i$ is the fading channel between the RIS $i^{th}$ reflecting surfaces to a single-antenna FU. $n_{fu}$ is AWGN with zero-mean and variance $N_0/2$ .

The transmit signal from the BS and the received signal at both NU and FU receivers is composed
of a superposition of the transmit signals of both the users. NU uses SIC to detect and decode its own data by subtracting FU data. On the contrary, FU only requires conventional detector since it has more allocated power.

\subsection{BER Performance}
 Bit Error Rate (BER) performance of the proposed system is analyzed in this section using Moment-Generating Function (MGF) of the fading channel distribution \cite{simon2005digital}.

The instantaneous SNR at NU is given by
\begin{equation}
\gamma_{inst(nu)}=\frac{\left| \sum_{i=1}^{N/2} \alpha_{i} \beta_{i} e^{j\left(\phi_{i}-\theta_{i}-\psi_{i}\right)}\right|^{2} E_{s}}{N_{0}}
\end{equation}

 As RIS is configured such that $\phi_{i}=\theta_{i}+\psi_{i}$ for $i~=~1,\ldots ,~N/2$,
the maximum instantaneous SNR is achieved and it is given by :
\begin{equation}
\gamma_{inst(nu)}=\frac{\left|\sum_{i=1}^{N/2} \alpha_{i} \beta_{i} \right|^{2} E_{s}}{N_{0}}
\end{equation}

Let $A$ be $\left|\sum_{i=1}^{N/2} \alpha_{i} \beta_{i} \right|$. Then it can be written as
\begin{equation}
\gamma_{inst(nu)}=\frac{A^{2} E_{s}}{N_{0}}
\end{equation}

Similarly for FU, the maximum instantaneous SNR is given by
\begin{equation}
\gamma_{inst(fu)}=\frac{\left|\sum_{i=(N/2) +1}^{N} \alpha_{i} \beta_{i} \right|^{2} E_{s}}{N_{0}} 
\end{equation}

Let $B$ be $\left|\sum_{i=(N/2) +1}^{N} \alpha_{i} \beta_{i} \right|$. Then it can be written as
\begin{equation}
\gamma_{inst(fu)}=\frac{B^{2} E_{s}}{N_{0}}
\end{equation}

Since $\alpha_{i}$ and $\beta_{i}$ are independently Rayleigh distributed RVs, the mean and the variance of their product are $E[\alpha_{i} ~ \beta_{i}] = \sigma^2 \left[\frac{\pi}{2}\right]$, $VAR[\alpha_{i} ~ \beta_{i}] =4\sigma^2 \left[1 - \frac{\pi^2}{16}\right]$ \cite{prodray}. Different values of $\sigma^2$ are alloted accordingly for NU and FU.  According to the Central Limit Theorem
(CLT), for a sufficiently large number of reflecting metasurfaces, i.e., $N >> 1$, $A$ and $B$ become Gaussian distributed random variables with parameters,
\begin{center}
$E[A] =  N_{nu} \sigma^2 \left[\frac{\pi}{2}\right]$ and VAR$[A]= 4N_{nu} \sigma^4 \left[1 - \frac{\pi^2}{16}\right]$\\
 \end{center}
$ N_{nu}$ and $N_{fu}$ are the arbitrary number of reflector elements allocated for NU and FU respectively. Similarly VAR$[B]$ is calculated. Now $\gamma$ is a non-central chi-square random variable with one degree of freedom and has the MGF 
$M(s ; \lambda)=\sqrt{(1-2 s)}{\exp \left(\frac{\lambda s}{1-2 s}\right)}$, where $\lambda$ is square of expectation, $s$ is the moment of the MGF. For a generalized $N$, MGF of instantaneous SNR is given by,
\begin{equation}
M_{\gamma}(s)=\left(\frac{1}{1-\frac{s N\left(16-\pi^{2}\right) E_{s}}{8 N_{0}}}\right)^{\frac{1}{2}} \exp \left(\frac{\frac{s N^{2} \pi^{2} E_{s}}{16 N_{0}}}{1-\frac{s N\left(16-\pi^{2}\right) E_{s}}{8 N_{0}}}\right)
\label{MGF}
\end{equation}
From (\ref{MGF}), using the approach given in \cite{Craig}, we can compute the average error probability of M-ary Phase Shift Keying (M-PSK) signaling is expressed as,
\begin{equation}
\begin{aligned}
P_{e}=\frac{1}{\pi} \int_{0}^{\pi / 2}&\left(  \frac{1}{1+\frac{N\left(16-\pi^{2}\right) E_{s}}{8  N_{0} \sin ^{2} \xi}}\right)^{\frac{1}{2}}  \exp \left(\frac{-\frac{N^{2} \pi^{2} E_{s}}{16 N_{0} \sin ^{2} \xi }}{1+\frac{N\left(16-\pi^{2}\right) E_{s}}{8 N_{0} \sin ^{2} \xi }}\right) d \xi
\end{aligned}
\label{prevs}
\end{equation}
In (\ref{prevs}) the error probability is maximum at $\xi=\pi/2$. Hence, the upper bound of error probability is given by
\begin{equation}
P_{e} \leq \frac{1}{2} \left(\frac{1}{1+\frac{N\left(16-\pi^{2}\right) E_{s}}{8  N_{0}}}\right)^{\frac{1}{2}}  \exp \left(\frac{-\frac{N^{2} \pi^{2} E_{s}}{16  N_{0}}}{1+\frac{N\left(16-\pi^{2}\right) E_{s}}{8  N_{0}}}\right) 
\end{equation}

 BER performance of downlink NOMA is derived in \cite{kayakara} for Rayleigh channels. Due to adverse channel conditions at FU, BPSK modulation is chosen and QPSK modulation is chosen for NU owing to relatively better channel condition. The BER performance at FU in conventional power domain NOMA is given by 
\begin{equation}
\begin{aligned} P_{FU}(e)= \frac{1}{2}\left[Q\left(\frac{(\sqrt{\varepsilon_{2}}+\sqrt{\varepsilon_{1} / 2}) }{\sqrt{N_{0} / 2}}\right)\right. \left.+Q\left(\frac{(\sqrt{\varepsilon_{2}}-\sqrt{\varepsilon_{1} / 2}) }{\sqrt{N_{0} / 2}}\right)\right] \end{aligned}
\label{Fuqunxn}
\end{equation}
%
\begin{figure}
\includegraphics[width=\linewidth]{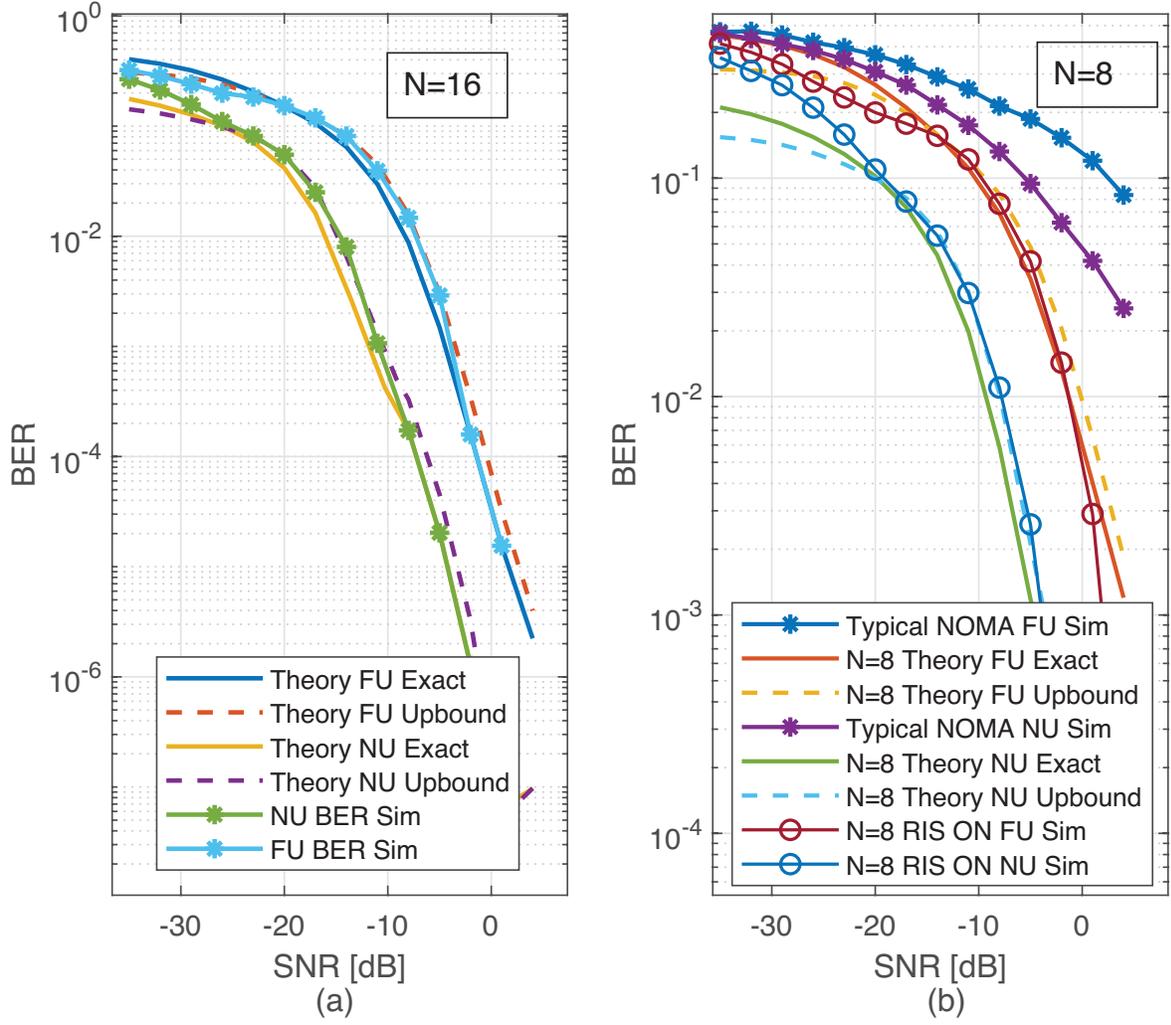}
\caption{ (a) BER Performance NOMA FU (BPSK) and NU (QPSK) $\alpha$=0.4, N=16. (b) BER Performance NOMA FU (BPSK) and NU (QPSK) $\alpha$=0.4, N=8 contrasted with conventional NOMA}
\label{N8}
\end{figure}
The error performance of SIC NU is given by
\begin{equation}
\begin{aligned}  P_{NU}\left(e \right)=  \frac{1}{4} \left[ Q\left(\sqrt{\frac{\varepsilon_{1}}{N_{0}}}\right) \times \left\{ 4-Q\left(\sqrt{\frac{{(\sqrt{2 \varepsilon_{2}}+\sqrt{\varepsilon_{1}})^{2}}}{N_{0}}}\right ) -Q\left(\sqrt{\frac{{(\sqrt{2 \varepsilon_{2}}-\sqrt{\varepsilon_{1}})^{2}}}{N_{0}}} \right) \right\}  -Q \left(\sqrt{\frac{{(\sqrt{2 \varepsilon_{2}}+\sqrt{\varepsilon_{1}})^{2}}}{N_{0}}} \right) \right]  \end{aligned}
\label{Nuqunxn}
\end{equation}

As always in any communication system with equiprobable symbols, most errors are contributed by signal points with lowest euclidean distances between them. As the sum and differences of the factors of $\varepsilon_{1}$ \& $\varepsilon_{2}$ has different unequal constellation of received NOMA symbol in all users. Replacing $E_s / N_0$ in (\ref{prevs}) (\ref{Nuqunxn}) by the NOMA symbol SNRs in (\ref{Fuqunxn}) , the BER performance at FU for the proposed RIS assisted downlink NOMA system is derived as

\begin{equation}
\begin{aligned}
P_{FU}\left(e \right)=\frac{1}{2\pi} &\left [\int_{0}^{\pi / 2}\left(\frac{1}{1+\frac{N_{fu}\left(16-\pi^{2}\right) (\sqrt{\varepsilon_{2}}+\sqrt{\varepsilon_{1} / 2})}{8 \sqrt{N_{0}} \sin ^{2} \xi}}\right)^{\frac{1}{2}} \exp \left(\frac{-\frac{N_{fu}^{2} \pi^{2} (\sqrt{\varepsilon_{2}}+\sqrt{\varepsilon_{1} / 2})}{16 \sqrt{N_{0}}\sin ^{2} \xi}}{1+\frac{N_{fu}\left(16-\pi^{2}\right) (\sqrt{\varepsilon_{2}}+\sqrt{\varepsilon_{1} / 2})}{8\sqrt{N_{0}} \sin ^{2} \xi }}\right) d \xi \right. + \\&
\int_{0}^{\pi / 2}\left(\frac{1}{1+\frac{N_{fu}\left(16-\pi^{2}\right) (\sqrt{\varepsilon_{2}}-\sqrt{\varepsilon_{1} / 2})}{8 \sqrt{N_{0}} \sin ^{2} \xi }}\right)^{\frac{1}{2}}\left. \exp \left(\frac{-\frac{N_{fu}^{2} \pi^{2} (\sqrt{\varepsilon_{2}}-\sqrt{\varepsilon_{1} / 2})}{16 \sqrt{N_{0}} \sin ^{2} \xi }}{1+\frac{N_{fu}\left(16-\pi^{2}\right) (\sqrt{\varepsilon_{2}}-\sqrt{\varepsilon_{1} / 2})}{8 \sqrt{N_{0}} \sin ^{2} \xi }}\right) d \xi 
\right]
\end{aligned}
\label{prev33sd33}%
\end{equation}
%

Similarly, the BER performance at NU of the proposed RIS assisted NOMA is given by

\begin{equation}
\begin{aligned}
P_{NU}\left(e \right)=\frac{1}{4\pi} &\left [\int_{0}^{\pi / 2}\left(\frac{1}{1+\frac{N_{nu}\left(16-\pi^{2}\right) ({\varepsilon_{1} })}{8\sqrt{N_{0}} \sin ^{2} \xi }}\right)^{\frac{1}{2}} \exp \left(\frac{-\frac{N_{nu}^{2} \pi^{2} ({\varepsilon_{1} })}{16 \sqrt{N_{0}} \sin ^{2} \xi }}{1+\frac{N_{nu}\left(16-\pi^{2}\right) ({\varepsilon_{1} })}{8 \sqrt{N_{0}} \sin ^{2} \xi }}\right) d \xi \hspace{0.5cm} \right.\times  \\&  
 \left\{ 4 -  \int_{0}^{\pi / 2}\left(\frac{1}{1+\frac{N_{nu}\left(16-\pi^{2}\right) (\sqrt{2\varepsilon_{2}}+\sqrt{\varepsilon_{1}})}{8 \sqrt{N_{0}} \sin ^{2} \xi }}\right)^{\frac{1}{2}} \exp \left(\frac{-\frac{N_{nu}^{2} \pi^{2} (\sqrt{2 \varepsilon_{2}}+\sqrt{\varepsilon_{1}})}{16  \sqrt{N_{0}} \sin ^{2} \xi}}{1+\frac{N_{nu}\left(16-\pi^{2}\right) (\sqrt{2\varepsilon_{2}}+\sqrt{\varepsilon_{1} })}{8 \sqrt{N_{0}} \sin ^{2} \xi }}\right) d \xi  \hspace{0.5cm} - \right.   \\&  \int_{0}^{\pi / 2} \left(\frac{1}{1+\frac{N_{nu}\left(16-\pi^{2}\right) (\sqrt{2\varepsilon_{2}}-\sqrt{\varepsilon_{1} })}{8 \sqrt{N_{0}} \sin ^{2} \xi }}\right)^{\frac{1}{2}} \left. \exp \left(\frac{-\frac{N_{nu}^{2} \pi^{2} (\sqrt{2 \varepsilon_{2}}-\sqrt{\varepsilon_{1}})}{16  \sqrt{N_{0}} \sin ^{2} \xi}}{1+\frac{N_{nu}\left(16-\pi^{2}\right) (\sqrt{2\varepsilon_{2}}-\sqrt{\varepsilon_{1} })}{8  \sqrt{N_{0}} \sin ^{2} \xi}}\right)^{\color{white}{1}} d \xi  \right\} \hspace{0.3cm} -\\& \int_{0}^{\pi / 2}\left(\frac{1}{1+\frac{N_{nu}\left(16-\pi^{2}\right) (\sqrt{2\varepsilon_{2}}+\sqrt{\varepsilon_{1}})}{8\sqrt{N_{0}} \sin ^{2} \xi }}\right)^{\frac{1}{2}}  \left. \exp \left(\frac{-\frac{N_{nu}^{2} \pi^{2} (\sqrt{2 \varepsilon_{2}}+\sqrt{\varepsilon_{1}})}{16\sqrt{N_{0}} \sin ^{2} \xi }}{1+\frac{N_{nu}\left(16-\pi^{2}\right) (\sqrt{2\varepsilon_{2}}+\sqrt{\varepsilon_{1} })}{8 \sqrt{N_{0}} \sin ^{2} \xi }}\right)^{\color{white}{1}} d \xi  \right]
 \end{aligned}
\label{prev666}
\end{equation}

From (\ref{prev33sd33}) and (\ref{prev666}), one cannot ignore any terms for approximation as the minimum distance pairs in superposition coding is dependent on $\alpha$. For example, a very low $\alpha$ gives better performance at FU but very poor performance at NU. 

\section{numerical and simulation results}

In this section, numerical and simulation results of BER performance of the proposed system is analyzed using (\ref{prev33sd33}) and (\ref{prev666}). Simulation is performed for BPSK for FU and QPSK for NU. To depict the real-time scenario FU channel is assumed to be having variance of $-3dB$ and NU channel is assumed to have $0dB$. 
 \subsection{ Effect of increasing reflecting elements $(N)$ }
\begin{figure}
\includegraphics[width=\linewidth]{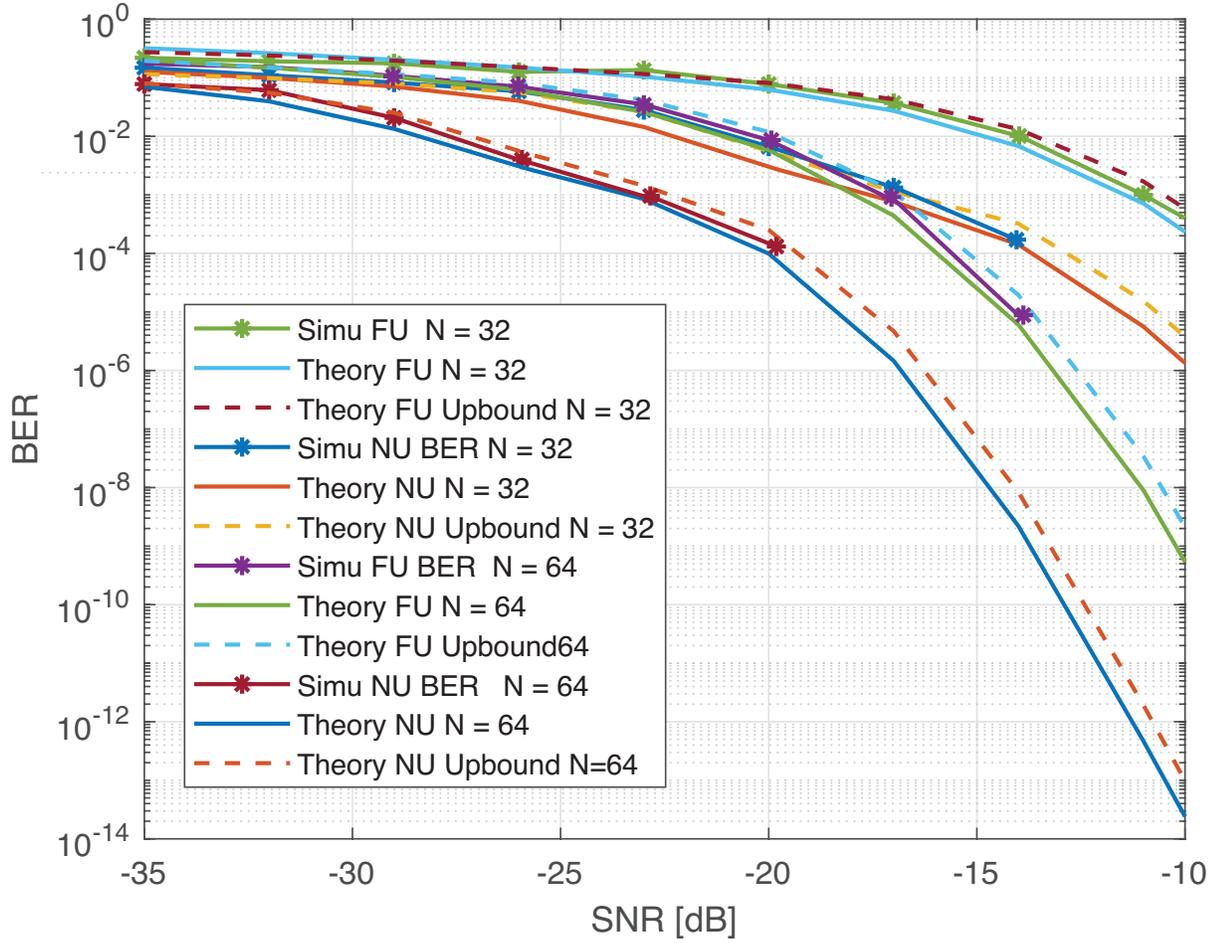}
\caption{ BER Performance NOMA FU BPSK and NU QPSK $\alpha=0.4, N= 32~\&~64 $}
\label{N32414}
\end{figure}

 The BER performance of the proposed system with $\alpha=0.4$, N=8 is shown in Fig.\ref{N8}(b) and for N=16 in Fig.\ref{N8}(a) respectively.
Here, $N_{nu}=N_{fu}=(N/2)$. Also when compared with typical NOMA in Fig.\ref{N8}(b), RIS assisted NOMA outperforms by good margin ($\approx 10dB$ for $N=8$). BER Performance of higher number of reflecting elements such as N=32 and 64 are illustrated in Fig.\ref{N32414}. As $N$ is doubled, gain is almost improved by $6dB$.
%
\subsection{Effect of $\alpha$ }
For a fixed $N=8$, NOMA superposition symbol power allocation factor $\alpha$ is varied from $0.1\rightarrow 0.4$ and the BER performance is depicted in Fig.\ref{N3244}(a). If $\alpha = 0.5$ it will cancel out symbols near the imaginary axis, if we increase $\alpha > 0.5$, it means more power allocation to NU, which is counter intuitive to user fairness. Therefore $\alpha \geq 0.5$ are ignored.
\subsection{Unequal RIS reflector allocation}
\begin{figure}
\includegraphics[width=\linewidth]{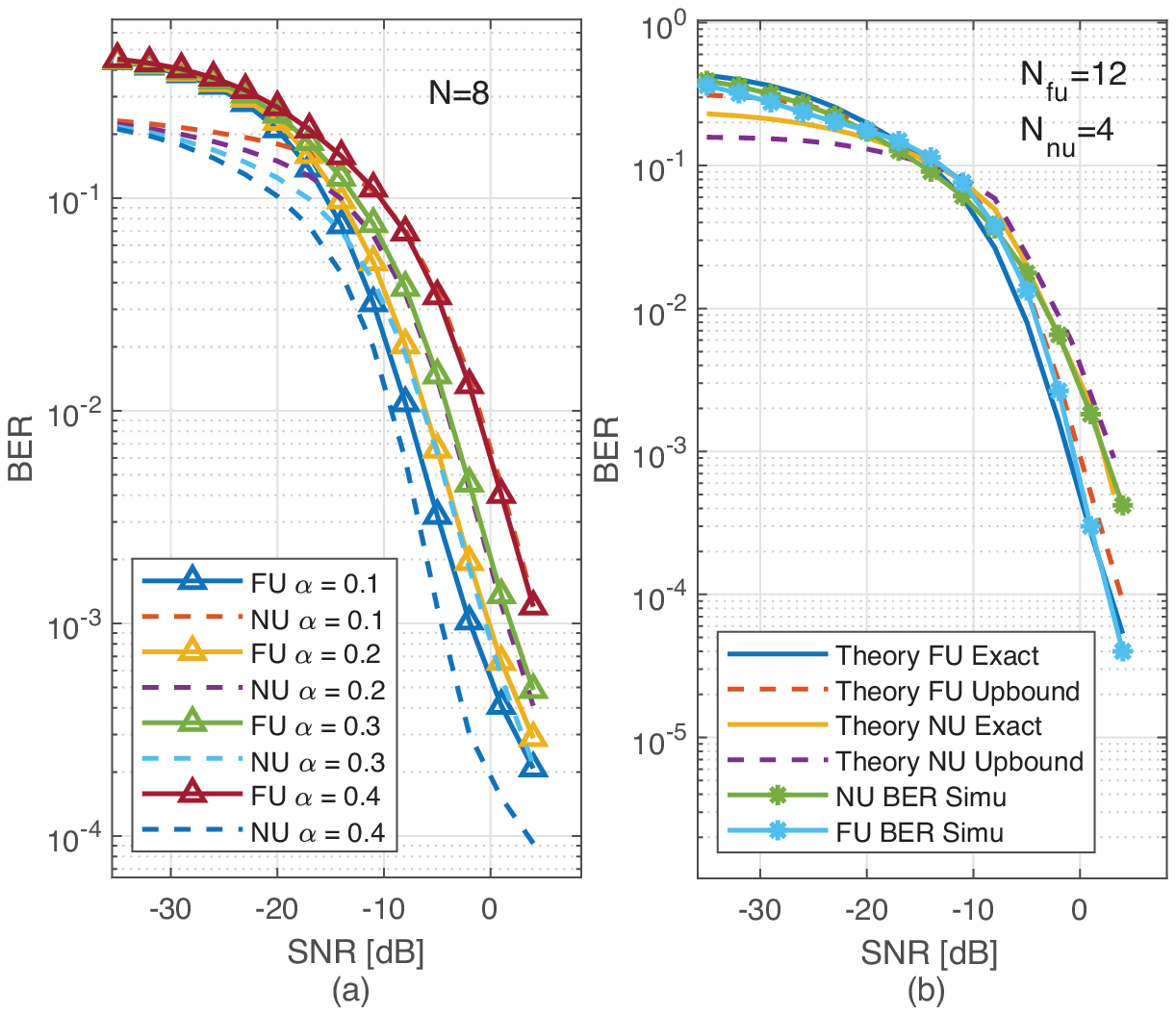}
\caption{(a) BER Performance NOMA FU BPSK and NU QPSK $\alpha=0.1\rightarrow 0.4$. (b) BER Performance of N= 16 NOMA FU $N_{FU}$=12 BPSK and NU $N_{NU}$=4 QPSK , $\alpha= 0.4$}
\label{N3244}
\end{figure}

To achieve fairness among all the NOMA users, one can unequally allocate the number of reflecting elements to get similar error performance for both FU and NU. In this way more reflectors can be used for users at a poorer channel conditions and fewer to those who already have a good link. The optimal selection of having more allotment to FU provides identical BER performance is shown in Fig.\ref{N3244}(b).

\section{Conclusion}

The gain achieved due to RIS is monumental, which argues the usecase in NOMA downlink even in the very low SNR to provide excellent reliability. RIS assisted NOMA downlink can accommodate even more users than previously capable and can also enable higher modulation schemes for most users. When $\alpha$ is varied and reliabilty is severely affected and one needs adaptive $\alpha$ control for practical NOMA systems based on the channel conditions. Similarly feedback based adaptive control of the reflecting elements allocation for NU and FU can be utilized to provide fairness in terms of error performance. The effect of M- QAM over multi-user scenarios and effect of RIS in generalised channels are being considered.


\end{document}